\documentclass[sigconf, nonacm=true, screen=true]{acmart}
\usepackage{xspace}
\usepackage{booktabs}
\usepackage{amsmath}
\usepackage{amsfonts}
\usepackage{array}
\usepackage{enumerate}
\usepackage{float}
\usepackage{subcaption}
\usepackage{graphicx}
\usepackage{multirow}
\usepackage{color}

\makeatletter
\g@addto@macro\bfseries{\boldmath}
\makeatother

\newcommand{\figref}[1]{Fig.~\ref{#1}}

\newcommand{\secref}[1]{Section~\ref{#1}}

\newcommand{\tabref}[1]{Table~\ref{#1}}

\usepackage{lipsum}

\usepackage{todonotes}

\renewcommand{\vec}[1]{\ensuremath{\mathbf{#1}}\xspace}
\newcommand{\p}{\vec{p}\xspace}
\newcommand{\q}{\vec{q}\xspace}

\newcommand{\MeV}{\text{Me\kern -0.15ex V}\xspace}
\newcommand{\GeV}{\text{Ge\kern -0.15ex V}\xspace}
\newcommand{\TeV}{\text{Te\kern -0.15ex V}\xspace}

\usepackage{bookmark}
\usepackage{algorithm}
\usepackage{lineno}
\usepackage{algorithmicx}
\usepackage{algpseudocode}
\usepackage{listings}
\usepackage{tabularx}
\usepackage{enumitem}
\AtBeginDocument{%
  \providecommand\BibTeX{{%
    \normalfont B\kern-0.5em{\scshape i\kern-0.25em b}\kern-0.8em\TeX}}}

\begin{document}

\title{\huge Grid-based minimization at scale: Feldman-Cousins corrections for SBN}

%
 
\author{Holger Schulz}
\affiliation{%
  \institution{Department of Physics, University of Cincinnati}
  \streetaddress{P.}
  \city{Cincinnati}
  \state{Ohio}
  \postcode{45219}
  \country{USA}
}
\author{Wes Ketchum}
\author{Jim Kowalkowski}
\author{Marc Paterno}
\author{Saba Sehrish}
\author{Marianette Wospakrik}
\affiliation{%
  \institution{Fermi National Accelerator Laboratory}
  \streetaddress{P.}
  \city{Batavia}
  \state{IL}
  \postcode{60510-0500}
  \country{USA}
}
\author{Guanqun Ge}
\author{Georgia Karagiorgi}
\author{Mark Ross-Lonergan}
\affiliation{%
    \institution{Department of Physics, Columbia University}
  \streetaddress{P.}
  \city{New York}
  \state{NY}
  \postcode{10027}
  \country{USA}
}
\renewcommand{\shortauthors}{}


\begin{abstract}
    We present a computational model for the construction of Feldman-Cousins
    (FC)~\cite{Feldman:1997qc} corrections frequently used in High Energy
    Physics (HEP) analysis. The program contains a grid-based minimization and
    is written in C++.  Our algorithms exploit vectorization through
    Eigen3~\cite{eigenweb}, yielding a  single-core speed-up of 350 compared to
    the original implementation, and achieve MPI data parallelism by using
    DIY~\cite{morozov_ldav16}. We demonstrate the application to scale very
    well at High Performance  Computing (HPC) sites. We use HDF5~\cite{hdf5} in
    conjunction with HighFive~\cite{highfive} to write results of the
    calculation to file.
\end{abstract}





\maketitle
\pagestyle{headings}
\section*{Conventions}
In this work, mathematical vectors are denoted as lower-case bold variables. Similarly,
we use upper-case bold variables to denote matrices. We are using the
Cori machines at the National Energy Research Scientific Computing
Center (NERSC)~\cite{corinersc}, and refer to them and their chips as Cori
phase 1 (Haswell) or Cori phase 2 (KNL). (See \tabref{tab:processors} for details.)

\begin{table}
    \centering
    \begin{tabular}{c|cc}
Machine        & Cori phase 1 (Haswell) & Cori phase 2 (KNL) \\ \toprule
CPU            & Intel Xeon E5-2698 v3 & Intel Xeon Phi 7250 \\
Clockspeed     & 2.3~GHz & 1.4~GHz \\
Cores per node & 32 & 68 \\ \bottomrule
    \end{tabular}
    \caption{Processors used in this work. }
    \label{tab:processors}
\end{table}

\section{Introduction}
\label{sec:introduction}

The Short Baseline Neutrino (SBN) 
experimental program~\cite{Antonello:2015lea}, comprising three 
Liquid Argon Time Projection Chamber (LArTPC) neutrino detectors, is
about to begin operations at Fermi National Accelerator Laboratory, in
order to study neutrino interactions from a
dedicated accelerator-based neutrino beam with unprecedented
precision. 
The experiment will measure and compare neutrino interaction rates recorded by
each detector to precisely test theories involving new particle
states, and typically involving a large number of theoretical model
parameters. Efforts are ongoing to determine the ``sensitivity'' of the
SBN experimental program to theoretical models, and further optimize
it in terms of a number of SBN experimental design parameters. 
Such sensitivity studies can be performed using the Feldman-Cousins
prescription~\cite{Feldman:1997qc}. 
This prescription provides a method for producing sensitivity contours
in a multi-dimensional parameter space (corrected frequentist
confidence intervals) through 
fits accounting for, generally, non-gaussian 
systematic uncertainties. The F.C.~prescription is computationally expensive and is 
typically implemented using multi-universe techniques.  One study run will typically 
use O(10M) CPU hours depending on the dimensionality of the problem and the 
desired resolution. The SBN collaboration employs the F.C.~analysis procedure 
as part of their analysis frameworks, including a framework known as
SBNfit~\cite{Cianci:2017okw}.  This paper provides a summary of the work we
have completed in adapting this SBNfit framework to run efficiently on
HPC facilities. This effort is particularly crucial for SBN, where
some of the theories that SBN aims to definitively test involve models
with up to $D=12$ independent fit parameters.

The task at hand is to perform a parameter scan in a $D$-dimensional parameter
space to find regions that are compatible with a central data vector $\vec{c}$
of length $N$, which represents the experimentally observed data.  Each point \p
is associated with a vector of length $N$, representing $N$
\emph{predicted} observations; we 
refer to this vector as signal or $\mathcal{S}(\p)$. The goal of the
scan is to build, through repeated fits to ``fake experimental data,'' a
distribution of a statistical measure ($\Delta\chi^2$) for each \p,
which effectively describes the number of degrees of freedom in the 
fit in the vicinity of that point. The effective number of degrees of freedom subsequently determine
associated critical 
values for the statistical measure used to quantify compatibility of
any given \p from a fit performed to ``real
experimental data''.

\begin{figure}[h]
    \centering
    \includegraphics[width=0.95\columnwidth]{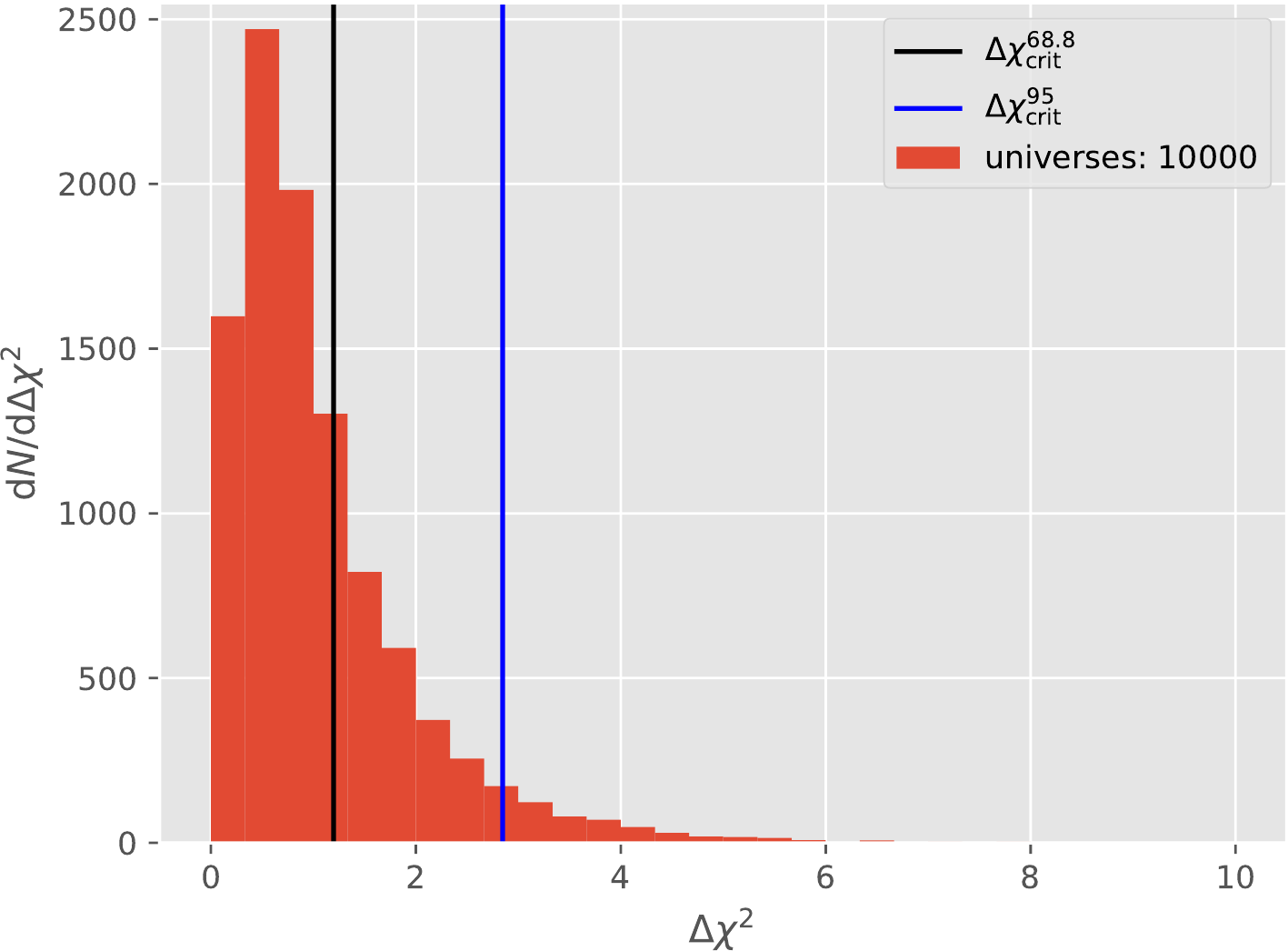}
    \caption{An example $\Delta\chi^2$ distribution for a single \p obtained
    from 10,000 universes. The vertical lines indicate the critical values of
$\Delta\chi^2$, left of which lie 68.8\% and 95\% of the distribution,
as indicated on the legend.}
    \label{fig:deltachi2}
\end{figure}

The procedure to build the
distribution of $\Delta\chi^2$ is performed using a multi-universe approach, 
where, in each universe, the input data is fluctuated around its mean
following established techniques. The number of universes to simulate is a
program parameter that we will refer to as $N_\text{univ}$.
An example distribution of $\Delta\chi^2$ along with an illustration of
critical values can be found in \figref{fig:deltachi2}.

At the heart of the computation is the calculation of a quantity called
$\chi^2$. The inputs are two vectors $\vec{a}$ and $\vec{b}$ of size $N$ and a
symmetric, positive definite matrix, $\vec{M}$, of size $N\times N$,
representing the inverese of a ``covariance matrix.'' The covariance matrix encapsulates
systematic and statistical uncertainties on the prediction vector.
The $\chi^2$ measure allows for a statistical interpretation of the numerical similarity of
$\vec{a}$ and $\vec{b}$ taking into account uncertainties and
correlations encoded in $\vec{M}$: 

\begin{equation}\label{eq:chi2}
    \chi^2 = (\vec{a} - \vec{b})^\top \cdot \vec{M} \cdot  (\vec{a} - \vec{b})
\end{equation}


The paper is organized as follows: in \secref{sec:algorithm} we introduce
the algorithm used for the computation of the Feldman-Cousins correction. We
follow in \secref{sec:hpc} with detailed discussions of challenges and
solutions to achieve computational performance and scalability. The achieved
performance and eventual limitations are explained in \secref{sec:performance}.
We make our concluding remarks in \secref{sec:conclusions} and discuss
prospects of future development.

\section{Algorithm}
\label{sec:algorithm}

The computational approach chosen for the parameter scan is to first discretize
the parameter space by means of a rigid rectangular grid, which is then linearized to take
on the form of a list ($\mathcal{G}$) of the points in the $D$-dimensional grid. We
will refer to the length of $\mathcal{G}$ as $N_\text{gridpoints}$.

To build the distribution $\Delta\chi^2$ for a single point $\p\in\mathcal{G}$,
a total of $N_\text{univ}$ calculations, comparing a given
(fluctuated) 
prediction at $\p$ with first-principles-predictions at all other
$N_\text{gridpoints}$ points, are performed. Algorithm~\ref{alg:deltaAchi} shows
a pseudo-code representation of the most relevant steps.  
For a single universe and following
the notation in \eqref{eq:chi2}, the procedure is as follows:

\begin{enumerate}
    \item Calculate signal prediction $\mathcal{S}(\p)$
    \item Fluctuate $\mathcal{S}(\p)$, yielding $\vec{d}$
    \item For each point $\q \in \mathcal{G}$: calculate $\chi^2(\vec{d}, \mathcal{S}(\q), \vec{M})$
    \item Determine $\q_\text{min}$ for which $\chi^2$ is minimal
    \item Update $\vec{M}$ for $\q_\text{min}$ and repeat scan over \q until
      $\chi^2_\text{min}$ convergence is reached at some new $\q_\text{min}$
    \item Return $\Delta\chi^2 = \chi^2(\vec{d}, \vec{c}, \vec{M}) - \chi^2(\vec{d}, \mathcal{S}(\q_\text{min}), \vec{M})$
\end{enumerate}


\noindent where \vec{c} is the central prediction (experimentally measured
data). This
means that the computational complexity, and therefore the program run time,
can be expected to scale $\propto N_\text{univ} \cdot N_\text{gridpoint}^2$.
Measurements of the scaling are given in more detail in \secref{sec:performance}.
An example distribution of $\Delta\chi^2$ along with an illustration of
critical values can be found in \figref{fig:deltachi2}. This procedure
follows the fit methodology of \cite{Aguilar-Arevalo:2013pmq} as a typical HEP implementation of the F.C.~correction.

\begin{algorithm}[!h]
    \caption{Building the $\Delta\chi^2$ distribution for a single $\p\in
    \mathcal{G}$ and a central prediction vector $\vec{c}$. In principle there
can be an additional loop around lines 7 to 14 aimed at updating the matrix
\vec{M}. We chose to omit this here for clarity.}
    \label{alg:deltaAchi}
    \begin{algorithmic}[1] 
        \Function{buildDeltaChi2}{$\p, \vec{c}, \vec{M}, \mathcal{S}, N_\text{univ} $}
        \State $\vec{s} = \mathcal{S(\p)}$ \Comment{prediction at \p}
        \For{$i\in[0,N_\text{univ})$}
            \State $\chi^2_\text{min}\gets \inf$;
            \State $\p_\text{min}=\p$
            \State $\vec{a} =$ fluctuation of $\vec{s}$
            \For{$\vec{q} \in \mathcal{G}$} \Comment{Iterate over all grid points}
                \State $\vec{b} = \mathcal{S(\vec{q})}$
                \State chi2=\textsc{calcChi2}$(\vec{a}, \vec{b}, \vec{M})$ \Comment{using Algorithm~\ref{alg:calcChi2}}
                \If{chi2 $< \chi^2_\text{min}$}
                    \State $\chi^2_\text{min}\gets$ chi2;
                    \State $\p_\text{min}=\vec{q}$
                \EndIf
            \EndFor
            \State deltaChi2 = \textsc{calcChi2}$(\vec{a}, \vec{c}, \vec{M})$ - $\chi^2_\text{min}$ 
            \EndFor
            \\The deltaChi2 values are aggregated and written to disk.
        \EndFunction
    \end{algorithmic}
\end{algorithm}


\begin{algorithm}
    \caption{$\chi^2$ calculation}
    \label{alg:calcChi2}
    \begin{algorithmic}[1] 
        \Function{calcChi2}{$\vec{a}, \vec{b}, \vec{M}$}
        \State $\vec{d}\gets\vec{a} - \vec{b}$
        \State $\chi^2=  \vec{d}^\top \cdot \vec{M} \cdot \vec{d}$
            \State \textbf{return} $\chi^2$
        \EndFunction
    \end{algorithmic}
\end{algorithm}

\section{Computational challenges and solutions}
\label{sec:hpc}

\begin{lstlisting}[caption={Implementation of \eqref{eq:chi2} using Eigen3},label={lst:chi2eigen}]
double calcChi(VectorXd const & a, 
               VectorXd const & b, 
               MatrixXd const & M )
{
   auto const & diff = a-b;
   return diff.transpose() * M * diff;
}
\end{lstlisting}

\begin{lstlisting}[caption={Original implementation of \eqref{eq:chi2}},label={lst:chi2orig}]
double calcChi(vector<double> a,
               vector<double> b, 
               vector<vector< double> > M, 
               int N)
{
    double chi2 = 0;
    for(int i =0; i<N; i++){
        for(int j =0; j<N; j++){
            chi2 += (a[i]-b[i]) * M[i][j] * (a[j]-b[j] );
        }
    }
    return chi2;
}
\end{lstlisting}

The calculation of $\Delta\chi^2$ at a single $p\in \mathcal{G}$ requires the
knowledge of the signal $\mathcal{S}(\p)$ for all $\p\in\mathcal{G}$. In the
original implementation the calculation of $\mathcal{S}(\p)$ was
computationally expensive due to the code not being vectorized, frequent access
of the file-system and the coming and going of objects. This is why the
original implementation resorted to precomputing and caching $\mathcal{S}(\p)$ ---
ultimately leading to severe limitations on the problem size due to memory
exhaustion. The re-implementation, which is heavily based on Eigen3, solves all
these issues and completely offsets\footnote{In fact, the reimplementation is already 350 times faster than the original code when run on a single core.} the performance gain from caching meaning
that the limitation on the size of problem that can be solved is solely due to
the available compute time. 

The computational complexity of the program is dependent on two factors. One being the 
number of multiverses, $N_\text{univ}$, that are simulated. We find the computational
cost to be linearly dependent on $N_\text{univ}$. The other factor being the number of
points, $N_\text{gridpoints}$ in the grid, which the run time depends on quadratically.

\begin{algorithm}
    \caption{Original implementation of computation of $\mathcal{S}(\p)$}
    \label{siggenorig}
    \begin{algorithmic}[1] 
        \Procedure{signal}{\p} \Comment{comment}
            \State Determine which files to open for \p
            \State Open 2 ROOT files and read TH1D objects into two vector<TH1D>
            \State Select physics parameters based on \p
            \State Call T1HD::Scale on elements of vector<TH1D> in a nested for-loop
            \State Compute a vector<double> from the vector<TH1D>
            \State Add this vector<double> to another vector<double>
            \State Return that vector<double>
        \EndProcedure
    \end{algorithmic}
\end{algorithm}


The major obstacle for an MPI-parallel version of the code that scales well on
an HPC system was the frequent access of the file system in the original
implementation. The same data was read from file any time objects were created.
This put an undue burden on the file system, ultimately leading to poor
performance and scaling. The solution was to read any of the input files only
once during the program life-time and to use collective MPI operations to copy
data to all ranks trough memory. In the original implementation, each
calculation of the $\mathcal{S}(\p)$ required the opening of two ROOT files
and the traversal and manipulation of the objects therein, which is very
expensive. Given that the $\mathcal{S}(\p)$ need to known for all $\p\in\mathcal{G}$
to compute the desired quantities, a caching of the $\mathcal{S}(\p)$
was present in the original implementation. Albeit improving the performance
of the program significantly, this approach ultimately lead to a restriction
on the problem size that could be tackled as the amount of memory needed
grows with $N_\text{gridpoints}$, ultimately exhausting the system memory.

Further, the coming and going of ROOT objects, such as TH1D histograms, turned
out to be a major performance bottleneck.  These ROOT histogram objects have many mathematical operations associated with them.  Performance issues were eliminated by   substituting these histograms objects with 
data types and operations from Eigen3. The histogram methods turned out to be a non-optimal
choice for necessary mathematical manipulations in this program, as the bulk
of the computations are matrix-vector multiplications and array operations. 
Eigen3's VectorXd and MatrixXd are simply much better suited for these operations, as
they inherently make use of the vectorisation and thus grant a
significant performance benefit --- so much so that the original gain from
caching the computations of the $\mathcal{S}(\p)$ in memory could completely be
offset. A comparison of signal prediction rates is shown in \tabref{tab:freqsiggen}.
These changes effectively transform the program from one limited by available memory into a one
limited only by CPU.

\begin{table}
    \centering
    \begin{tabular}{l|l}
        & Frequency of $\mathcal{S}(\p)$ \\ \toprule
        Original, Haswell & 30~Hz \\
        New, Haswell & 10~MHz \\
        New, KNL & 2~MHz \\ \bottomrule
    \end{tabular}
    \caption{Comparison of single-core signal prediction rates. The achieved speed-up allows to avoid having to cache $\mathcal{S}(\p)$ and thus eliminate the memory-limitation of the problem size.}
    \label{tab:freqsiggen}
\end{table}

A further important change was the use of HDF5 as output file format.  We are
using the MPI-capable file objects for writing to disk which reduces necessary
MPI communications to a minimum. This is especially suitable for the problem at
hand as the information written do disk is comprised of trivial data types.
Further, the amount of data written by every rank as well as their target location
within the HDF5 datasets is is completely deterministic and thus known upon
start of the program.

The final revised application utilizes DIY\cite{morozov_ldav16} for data parallelism across processes and nodes, and Eigen3 for linear algebra and vectorized calculations.  The application is now parameterized on number of universes, the number of grid points, and the standard MPI option for number of ranks.

\section{Performance and scaling}
\label{sec:performance}

In this section we measure the overall program execution time as a function of
$N_\text{univ}$ and $N_\text{gridpoint}$ to test our initial hypothesis that
the run time should scale $\propto N_\text{univ} \cdot N_\text{gridpoint}^2$.
We further demonstrate strong scaling of the application and make an attempt at
predicting the maximum size problem that can be solved when running on all of
NERSC's Cori for a full day.

\subsection{Scaling with $N_\text{univ}$ and $N_\text{gridpoint}$}

\begin{figure}[h]
    \centering
    \includegraphics[width=\columnwidth]{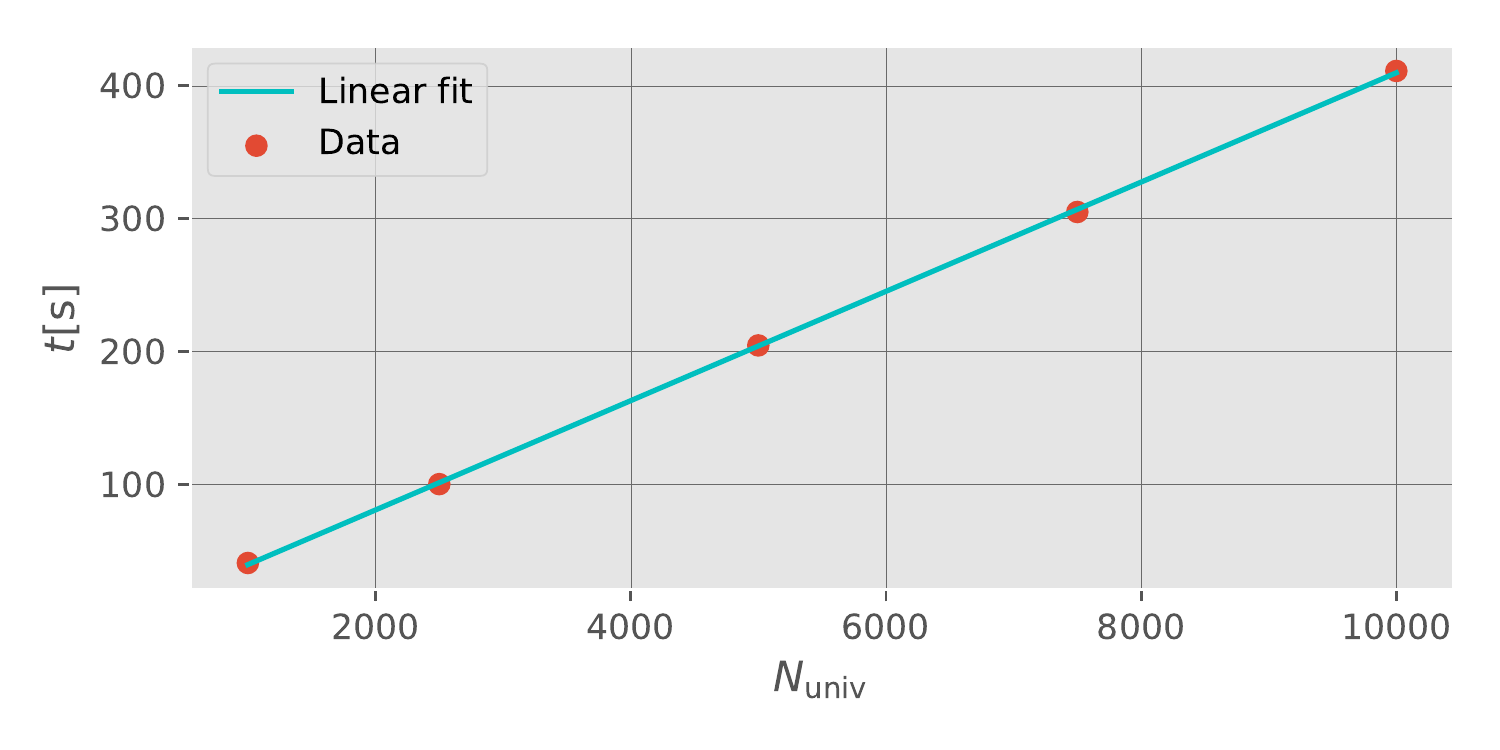}
    \caption{Scaling of the program run time with the number of universes, demonstrating a linear dependence on $N_\text{univ}$.}
    \label{fig:scalingnuniv}
\end{figure}

To collect the data, we used a fixed number of Haswell nodes (and therefore
ranks) and varied $N_\text{univ}$ and $N_\text{gridpoint}$ independently.
\figref{fig:scalingnuniv} show the scaling of the program time for a fixed
number $N_\text{gridpoint}$ as a function of $N_\text{univ}$. The data is
overlayed with a linear fit and there is no visible deviation from a linear
scaling.

\begin{figure}[h]
    \centering
    \includegraphics[width=\columnwidth]{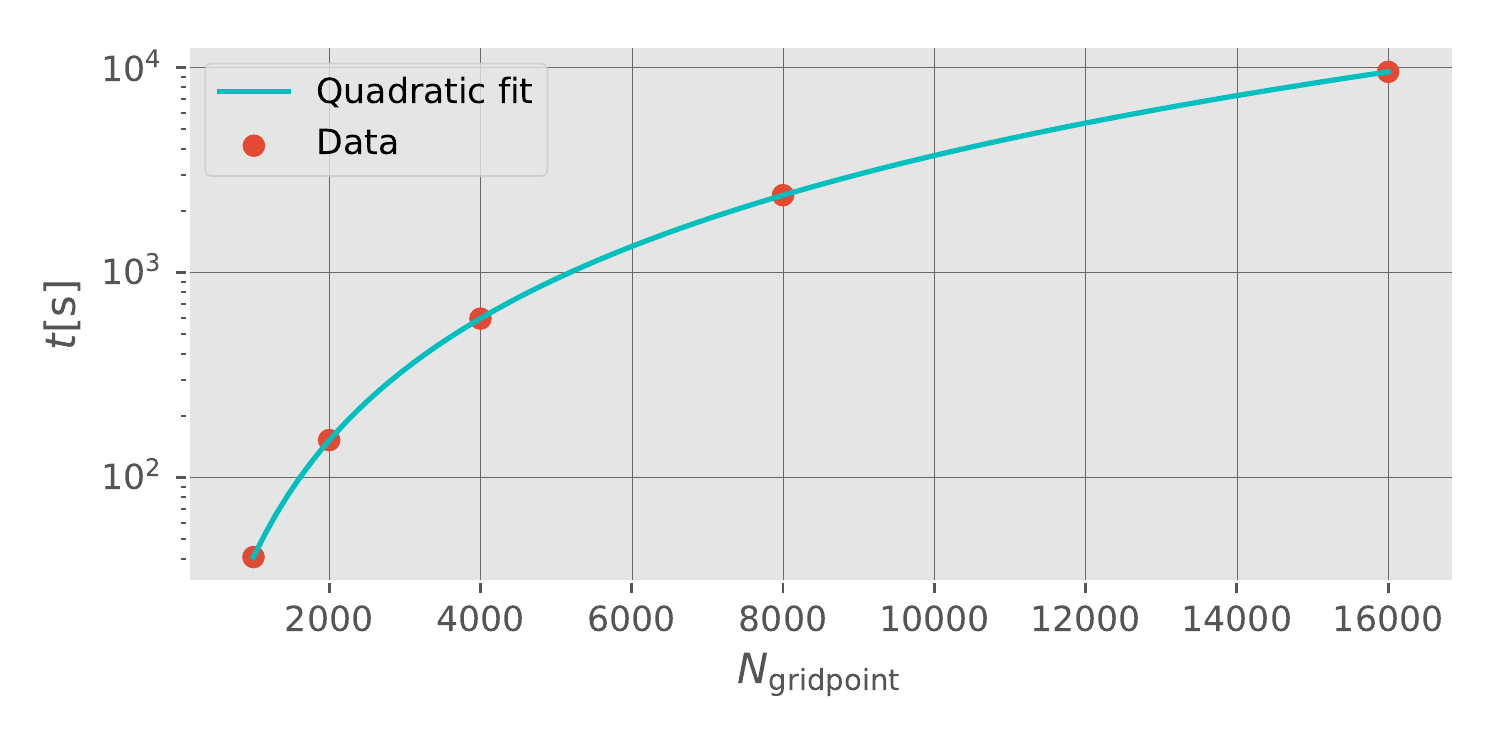}
    \caption{Scaling of the program run time with the number of grid points, demonstrating a quadratic dependence on $N_\text{gridpoint}$.}
    \label{fig:scalingngridp}
\end{figure}

Similarly, in \figref{fig:scalingngridp} we fix the number of universes and
measure the program execution time as function of $N_\text{gridpoint}$. The
quadratic fit to the data again shows no visible deviation from the expected
behavior.

\subsection{Single node scaling}

We measure the program execution time as a function of the number of ranks in order to
see in how far the program benefits from using multiple MPI ranks. 
Haswell nodes allow to run
on up to two ranks per core, leading to a maximum of 64 ranks per node. 
On the KNL nodes it is possible to utilize up to four hardware threads per processor, 
resulting in a maximum
number of ranks of 272 per node (See also \tabref{tab:processors}).

\begin{figure}
    \centering
    \includegraphics[width=\columnwidth]{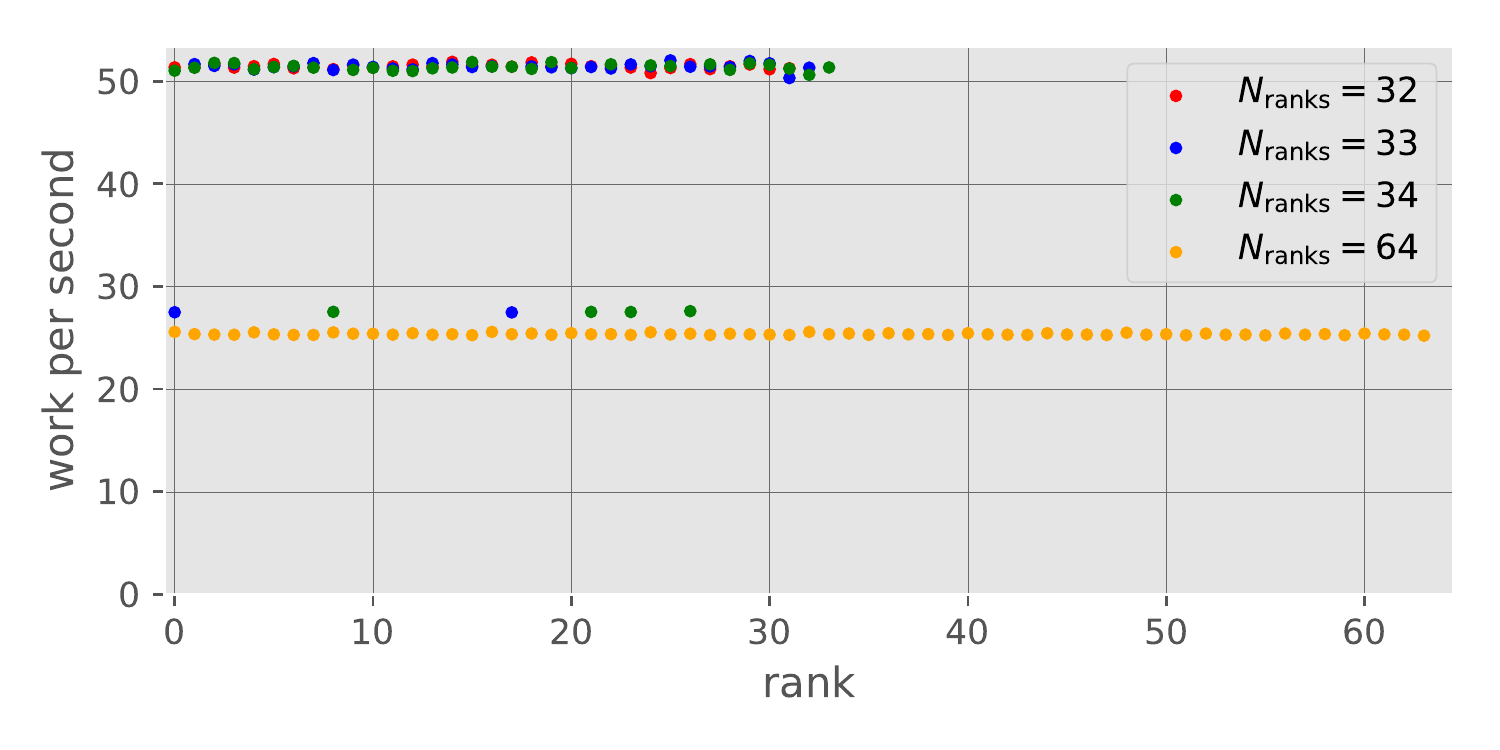}
    \caption{Measurement of the single Haswell node performance for a fixed problem size. The work is distributed among ranks and we measure the utilization of each rank in terms of units of work per second. Non-integer multiples of the number of cores available leads to a significant performance loss. We do not observe a benefit from using 2 ranks per core.}
    \label{fig:threadshaswell}
\end{figure}

\begin{figure}
    \centering
    \includegraphics[width=\columnwidth]{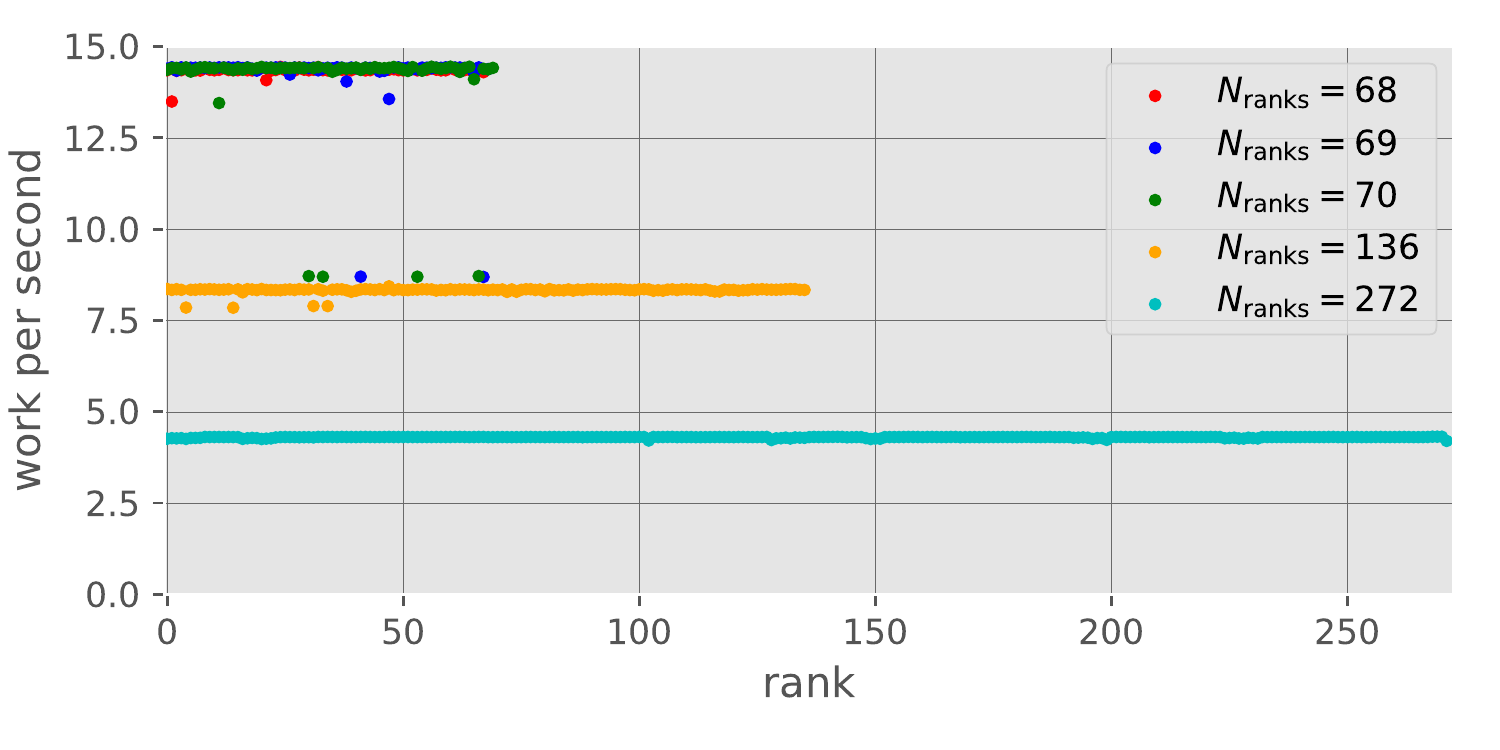}
    \caption{Measurement of the single KNL node performance for a fixed problem size. The work is distributed among ranks and we measure the utilization of each rank in terms of units of work per second. Non-integer multiples of the number of cores available leads to a significant performance loss. We observe a mild benefit from using 2 and 4 ranks per core.}
    \label{fig:threadsknl}
\end{figure}

We select a fixed size problem, distribute the work among the ranks as evenly
as possible, and record the time spent in the main part of the program for each
rank separately. 
For this application, the work is number of points from the parameter space and number of universes to generate.  
The data is shown in \figref{fig:threadshaswell} for Haswell
nodes and in \figref{fig:threadsknl} for KNL. The most relevant quantity in
terms of throughput is of course the amount of time the slowest rank spends
($t_\text{max}$). For Haswell nodes we find the best performance when using 32
ranks, meaning our program does not benefit from threading. We also find that
when using e.g. 33 or 34 ranks, i.e.  effectively oversubscribing 1 and 2 cores
respectively, the performance in terms of the total run time drops almost by a
factor of two. The effect can be seen in \figref{fig:threadshaswell}. When
utilizing 33 ranks, we are in a situation where two ranks are competing for
resources on the same core.  Similarly, when running with 34 ranks, we observe
4 ranks that are slower than the rest since they are visibly competing with
each other. We repeat the measurements on a KNL node and find a similar
picture. Again, using non-integer multiples of the number of cores (68) leads
to an overall worse performance due to ranks competing for resources in an
unbalanced way. In contrast to Haswell, however, we observe a gain in
performance in terms of $t_\text{max}$ from using 2 and 4 ranks per core of
16\% and 24\% respectively.

\begin{table}[!h]
    \centering
    \begin{tabular}{c|ccc}
        Ranks                & 68 & 136 & 272 \\ \toprule
        $t_\text{max}$[s]    & 87 & 75 & 70   \\ 
        Gain w.r.t. 68 ranks &    & 16\% & 24\% \\\bottomrule
    \end{tabular}
    \caption{Total run time of a fixed problem on a single KNL node demonstrating that there is a mild benefit in using hardware threading.}
    \label{tab:knlgain}
\end{table}

\subsection{Multi node scaling}

For any distributed computing program it is crucial to know whether using N times as many
computing resources yields a program execution time that is N times faster. In order to
study this, we define a reasonable large problem of a fixed size and measure the time it
takes to complete the main part of the program as a function of the number of Haswell nodes
used ($N_\text{nodes}$). Following the finding in the previous section, we use 32 ranks per node in this study.

We further define the program execution efficiency as 
\begin{equation}\label{eq:efficiency}
    \varepsilon = N_\text{nodes}\cdot\frac{t(N_\text{nodes})}{t(1)},
\end{equation}
where $t(1)$ is the run time on a single.

\begin{figure}[h]
    \centering
    \includegraphics[width=\columnwidth]{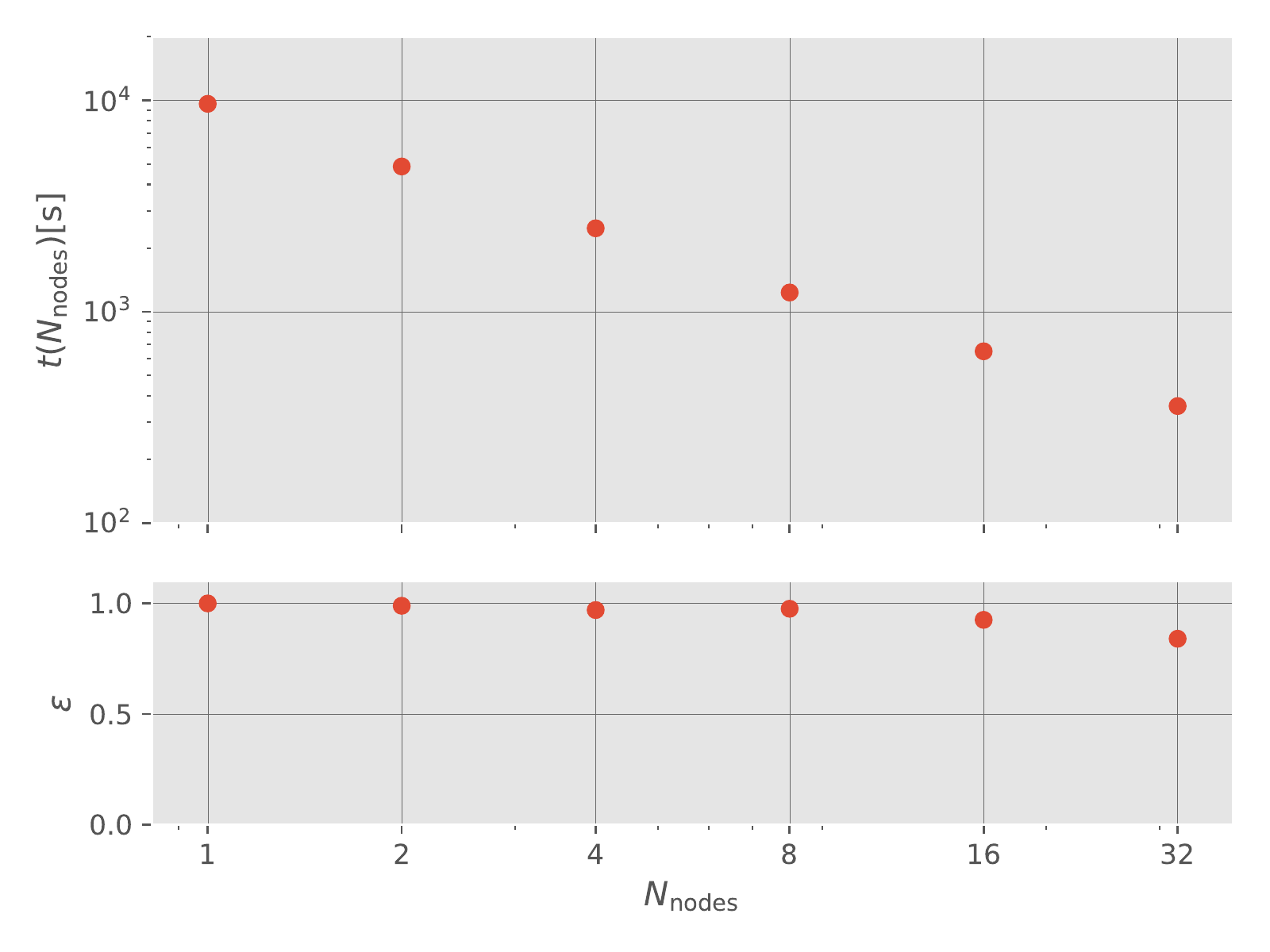}
    \caption{Strong scaling measurements on Haswell nodes. We observe generally good scaling up to the point where the
    amount of work per rank becomes relatively small.}
    \label{fig:strongscaling}
\end{figure}

Our measurements are shown in \figref{fig:strongscaling} and demonstrate that the
program scales reasonably well. Eventually the amount of work to be done per rank
becomes relatively small and the drop to about 80\% efficiency at 32 nodes (1024 ranks)
indicates that the compute time of individual tasks may not be completely uniform.

\subsection{Estimation of what is possible}

\begin{figure}
    \centering
    \includegraphics[width=\columnwidth]{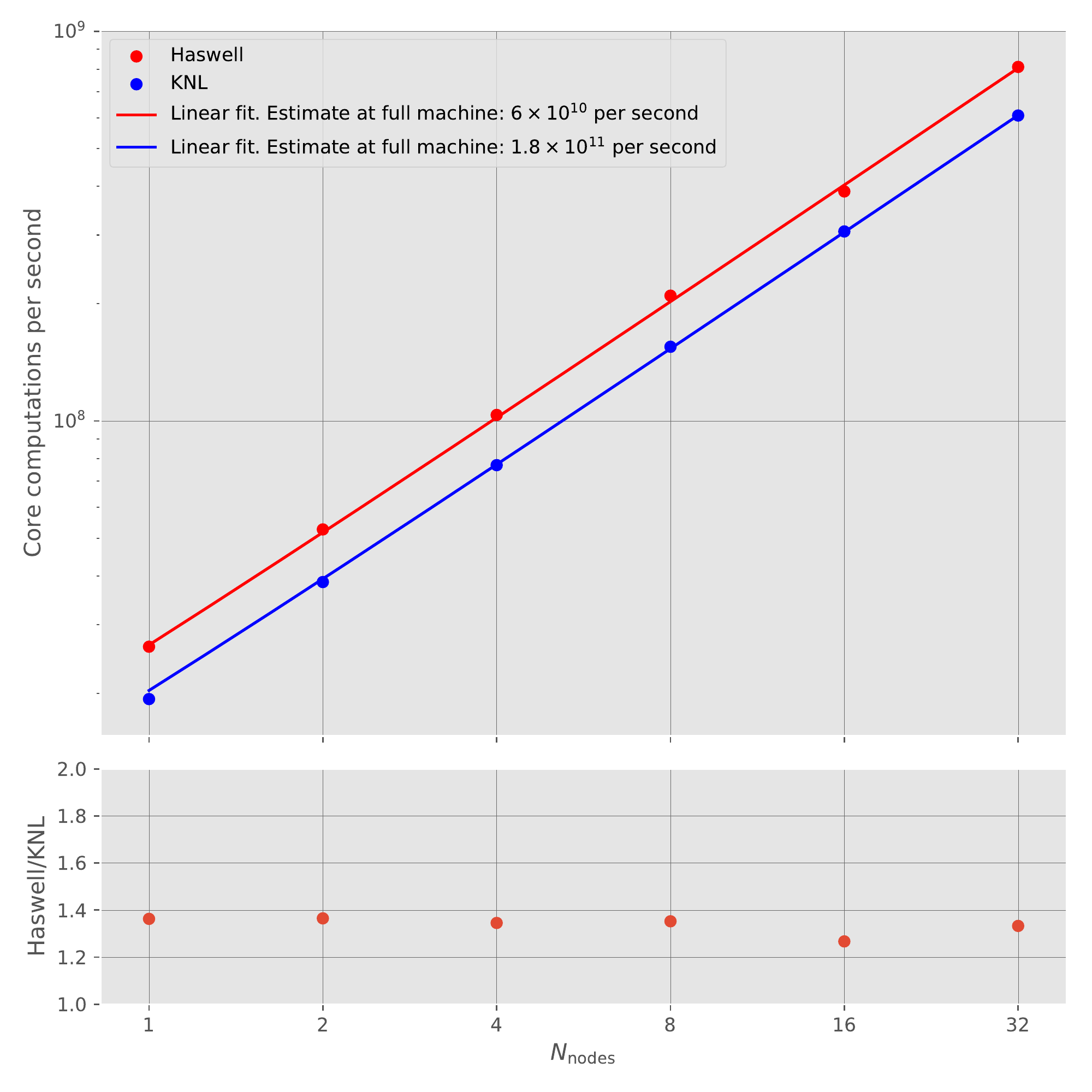}
    \caption{``Velocity'' plot showing the ``speed'' of core computations on different architectures as function of the number of nodes.}
    \label{fig:velocity}
\end{figure}

In order to estimate the size of problem that can be tackled with a machine
like Cori, we define the computation of $\mathcal{S(\vec{q})}$ and successive
execution of {\tt calcChi} (lines 8 and 9 of Algorithm~\ref{alg:deltaAchi}) as
\emph{core computation} as it is a quantity that is independent of
$N_\text{gridpoint}$ and $N_\text{univ}$.  We then measure the numbers of
\emph{core computations} per second as a function of the number of nodes. We
find the scaling to be linear in the number of nodes for our measurements. This
allows to estimate an upper limit of of \emph{core computations} if the whole
machine were available. For Cori phase 1 (Haswell) we estimate an upper
boundary of $6\time10^{10}~s{-1}$ from a linear extrapolation to the entirety
of 2388 nodes. Similarly, we find an upper boundary of $1.8\time10^{11}~s{-1}$
when using all 9688 nodes of Cori phase 2 (KNL). The data and linear fits
are shown in \figref{fig:velocity}. It should be noted that the linear
scaling assumption should be considered optimistic and be interpreted as
an upper limit of the performance of the program. 

We can use these numbers to estimate the size of, $\mathcal{G}$, that
could possibly be solved with our implementation of the Feldman-Cousins method.
A commonly occurring measure of statistical accuracy in that method is a
significance level expressed in multiples of $\sigma$. For instance, $3\sigma$
and $5\sigma$ correspond to $N_\text{univ}=10^4$ and $N_\text{univ}=10^8$,
respectively. In \tabref{tab:bounds} we provide numbers of maximally
possible grid sizes under the assumption of running on the entirety of
Cori phase 1/2 for a full day. We find that for the $5\sigma$ significance
level, which is typically applied to claim scientific discovery, only
comparatively small problem sizes of about 10 thousand grid points can be solved.
This translates into two-dimensional problems of fairly good resolution of
100 points per axis. In three dimensions, the resolution drops to
about 20 points per axis. For higher dimensions the application of this
method becomes computationally intractable.

\begin{table}
    \centering
    \begin{tabular}{l|c|c}
                                & $N_\text{univ}=10^4$ ($3\sigma$) & $N_\text{univ}=10^8$ ($5\sigma$)\\\toprule
        Cori phase 1 (Haswell)  & $7.2\times 10^{5}$ & $7.2\times 10^{3}$ \\
        Cori phase 2 (KNL)      & $1.2\times 10^{6}$ & $1.2\times 10^{4}$ \\\bottomrule
    \end{tabular}
    \caption{Upper boundaries on grid sizes that can be processed when running \emph{a full day} on all of  Cori phase 1/2.}
    \label{tab:bounds}
\end{table}

\section{Conclusions}
\label{sec:conclusions}

The goal of this work has been to reconstruct the existing SBN application 
and algorithms that calculate Feldman-Cousins corrections to run efficiently
on current state-of-the-art processor architectures and systems available at HPC facilities.
We transformed a High Energy Physics problem common within neutrino physics 
from a serial-execution program limited by machine memory into an MPI-parallel 
application that scales up to available compute power of a facility.   
We obtain node-level and thread-level parallelism through DIY, 
and obtain significant performance improvements by restructuring algorithms 
to use Eigen3 for matrix multiplications and array manipulations.

An interesting side-effect of this work is the calculations performed within 
functions such as calcChi2 better represent the actual mathematical operations being 
performed on the data.  Eigen provides a higher-level abstraction that greatly
improves the readability of the code.   This will allow for physics-developers to
understand what is happening in this code if and when problems arise.  It also provides
better opportunities for the compiler and library to optimize the executable for
specific platforms and architectures.  This ought to be of great benefit to the experiment.
Since the algorithms and procedures in this analysis application are so similar to 
experiments other than SBN, we believe the techniques employed here will provide
similar benefits in terms of performance and design for the broader neutrino physics 
program.

\subsection{Future work}

For this round of improvements, we chose to stay with more traditional large-core
libraries and techniques for vectorization within a compute node.  
Our next phase will explore changes necessary for evaluating the performance 
of GPU accelerators to further reduce the program execution time.  
For this future work, we will target Kokkos as a replacement for Eigen3.

We find that in spite of the achieved improvements, the computational
complexity of the Feldman-Cousins approach coupled with SBN design choices 
for the current implementation severely limit the dimensionally of the problem 
that can be handled.  The current ``brute-force'' method of scanning a regular grid
becomes prohibitive after two or perhaps three dimensions.   
An alternative scanning approach will need to be found to move towards the seven or 
more dimensions that the experiment would like to probe, along with 
allowing $N_\text{univ}$ to be different for each $\p\in\mathcal{G}$,
such that large numbers of universes are applied only in the vicinity of the
eventual contour.   We will be exploring alternative techniques to this grid 
scan through our connection with the SciDAC FASTmath Institute~\cite{fastmath}. 

Incorporation of the new capabilities enabled by this work into the SBN experiment
analysis workflow is underway.  The reorganization of this application and efficient use of computational
resources opens the door to incorporating more dimensions into the
problem space that SBN aims to address. 
Replacement of the well-established and validated grid-scan procedure will require 
additional consideration and study within the SBN collaboration.  Providing an 
alternative procedure and comparison study is part of our future work; the physics
study and impact within SBN, however, falls outside of our scope. Nevertheless, our work
ought to provide the machinery to greatly improve, for example, design
parameters such as how much data will be necessary for the experiment to 
collect in order to achieve any given future physics goals.



\begin{acks}
    This material is based upon work supported by the U.S. Department of
    Energy, Office of Science, Office of Advanced Scientific Computing
    Research, Scientific Discovery through Advanced Computing (SciDAC) program,
    grant ``HEP Data Analytics on HPC'', No.~1013935.  
    It was supported by the U.S. Department of Energy under contracts
    DE-AC02-76SF00515 and DE-AC02-07CH11359 and used resources of the National
    Energy Research Scientific Computing Center (NERSC), a U.S.  Department of
    Energy Office of Science User Facility operated under Contract No.
    DE-AC02-05CH11231. GG, GK and MRL are supported by the National Science Foundation under Grant No. PHY-1707971.
\end{acks}

\bibliographystyle{ACM-Reference-Format}
\bibliography{fchephpcpaper.bib}


\begin{thebibliography}{10}


\ifx \showCODEN    \undefined \def \showCODEN     #1{\unskip}     \fi
\ifx \showDOI      \undefined \def \showDOI       #1{#1}\fi
\ifx \showISBNx    \undefined \def \showISBNx     #1{\unskip}     \fi
\ifx \showISBNxiii \undefined \def \showISBNxiii  #1{\unskip}     \fi
\ifx \showISSN     \undefined \def \showISSN      #1{\unskip}     \fi
\ifx \showLCCN     \undefined \def \showLCCN      #1{\unskip}     \fi
\ifx \shownote     \undefined \def \shownote      #1{#1}          \fi
\ifx \showarticletitle \undefined \def \showarticletitle #1{#1}   \fi
\ifx \showURL      \undefined \def \showURL       {\relax}        \fi
\providecommand\bibfield[2]{#2}
\providecommand\bibinfo[2]{#2}
\providecommand\natexlab[1]{#1}
\providecommand\showeprint[2][]{arXiv:#2}

\bibitem[\protect\citeauthoryear{Aguilar-Arevalo et~al\mbox{.}}{Aguilar-Arevalo
  et~al\mbox{.}}{2013}]%
        {Aguilar-Arevalo:2013pmq}
\bibfield{author}{\bibinfo{person}{A.~A. Aguilar-Arevalo} {et~al\mbox{.}}}
  \bibinfo{year}{2013}\natexlab{}.
\newblock \showarticletitle{{Improved Search for $\bar \nu_\mu \rightarrow \bar
  \nu_e$ Oscillations in the MiniBooNE Experiment}}.
\newblock \bibinfo{journal}{\emph{Phys. Rev. Lett.}}  \bibinfo{volume}{110}
  (\bibinfo{year}{2013}), \bibinfo{pages}{161801}.
\newblock
\urldef\tempurl%
\url{https://doi.org/10.1103/PhysRevLett.110.161801}
\showDOI{\tempurl}
\showeprint[arxiv]{hep-ex/1303.2588}


\bibitem[\protect\citeauthoryear{Antonello et~al\mbox{.}}{Antonello
  et~al\mbox{.}}{2015}]%
        {Antonello:2015lea}
\bibfield{author}{\bibinfo{person}{M. Antonello} {et~al\mbox{.}}}
  \bibinfo{year}{2015}\natexlab{}.
\newblock \showarticletitle{{A Proposal for a Three Detector Short-Baseline
  Neutrino Oscillation Program in the Fermilab Booster Neutrino Beam}}.
\newblock  (\bibinfo{year}{2015}).
\newblock
\showeprint[arxiv]{physics.ins-det/1503.01520}


\bibitem[\protect\citeauthoryear{Cianci, Furmanski, Karagiorgi, and
  Ross-Lonergan}{Cianci et~al\mbox{.}}{2017}]%
        {Cianci:2017okw}
\bibfield{author}{\bibinfo{person}{Davio Cianci}, \bibinfo{person}{Andy
  Furmanski}, \bibinfo{person}{Georgia Karagiorgi}, {and} \bibinfo{person}{Mark
  Ross-Lonergan}.} \bibinfo{year}{2017}\natexlab{}.
\newblock \showarticletitle{{Prospects of Light Sterile Neutrino Oscillation
  and CP Violation Searches at the Fermilab Short Baseline Neutrino Facility}}.
\newblock \bibinfo{journal}{\emph{Phys. Rev.}} \bibinfo{volume}{D96},
  \bibinfo{number}{5} (\bibinfo{year}{2017}), \bibinfo{pages}{055001}.
\newblock
\urldef\tempurl%
\url{https://doi.org/10.1103/PhysRevD.96.055001}
\showDOI{\tempurl}
\showeprint[arxiv]{hep-ph/1702.01758}


\bibitem[\protect\citeauthoryear{{FASTmath Institute}}{{FASTmath
  Institute}}{[n.d.]}]%
        {fastmath}
\bibfield{author}{\bibinfo{person}{{FASTmath Institute}}.}
  \bibinfo{year}{[n.d.]}\natexlab{}.
\newblock \bibinfo{booktitle}{\emph{{FASTmath is a program of the Office of
  Science within the Department of Energy}}}.
\newblock
\newblock
\shownote{\url{https://fastmath-scidac.llnl.gov/}.}


\bibitem[\protect\citeauthoryear{Feldman and Cousins}{Feldman and
  Cousins}{1998}]%
        {Feldman:1997qc}
\bibfield{author}{\bibinfo{person}{Gary~J. Feldman} {and}
  \bibinfo{person}{Robert~D. Cousins}.} \bibinfo{year}{1998}\natexlab{}.
\newblock \showarticletitle{{A Unified approach to the classical statistical
  analysis of small signals}}.
\newblock \bibinfo{journal}{\emph{Phys. Rev.}}  \bibinfo{volume}{D57}
  (\bibinfo{year}{1998}), \bibinfo{pages}{3873--3889}.
\newblock
\urldef\tempurl%
\url{https://doi.org/10.1103/PhysRevD.57.3873}
\showDOI{\tempurl}
\showeprint[arxiv]{physics.data-an/physics/9711021}


\bibitem[\protect\citeauthoryear{Guennebaud, Jacob, et~al\mbox{.}}{Guennebaud
  et~al\mbox{.}}{2010}]%
        {eigenweb}
\bibfield{author}{\bibinfo{person}{Ga\"{e}l Guennebaud},
  \bibinfo{person}{Beno\^{i}t Jacob}, {et~al\mbox{.}}}
  \bibinfo{year}{2010}\natexlab{}.
\newblock \bibinfo{title}{Eigen v3}.
\newblock \bibinfo{howpublished}{\url{http://eigen.tuxfamily.org}}.
\newblock


\bibitem[\protect\citeauthoryear{Morozov and Peterka}{Morozov and
  Peterka}{2016}]%
        {morozov_ldav16}
\bibfield{author}{\bibinfo{person}{Dmitriy Morozov} {and} \bibinfo{person}{Tom
  Peterka}.} \bibinfo{year}{2016}\natexlab{}.
\newblock \showarticletitle{Block-Parallel Data Analysis with DIY2}.
\newblock  (\bibinfo{year}{2016}).
\newblock
\newblock
\shownote{\url{https://github.com/diatomic/diy}.}


\bibitem[\protect\citeauthoryear{{National Energy Research Scientific Computing
  Center}}{{National Energy Research Scientific Computing Center}}{[n.d.]}]%
        {corinersc}
\bibfield{author}{\bibinfo{person}{{National Energy Research Scientific
  Computing Center}}.} \bibinfo{year}{[n.d.]}\natexlab{}.
\newblock \bibinfo{booktitle}{\emph{The Cori System at the NERSC}}.
\newblock
\newblock
\shownote{\url{https://docs.nersc.gov/systems/cori/}.}


\bibitem[\protect\citeauthoryear{{The Blue Brain Project}}{{The Blue Brain
  Project}}{NNNN}]%
        {highfive}
\bibfield{author}{\bibinfo{person}{{The Blue Brain Project}}.}
  \bibinfo{year}{2016-NNNN}\natexlab{}.
\newblock \bibinfo{booktitle}{\emph{{HighFive}}}.
\newblock
\newblock
\shownote{\url{https://github.com/BlueBrain/HighFive}.}


\bibitem[\protect\citeauthoryear{{The HDF Group}}{{The HDF Group}}{NNNN}]%
        {hdf5}
\bibfield{author}{\bibinfo{person}{{The HDF Group}}.}
  \bibinfo{year}{1997-NNNN}\natexlab{}.
\newblock \bibinfo{booktitle}{\emph{{Hierarchical Data Format, version 5}}}.
\newblock
\newblock
\shownote{\url{http://www.hdfgroup.org/HDF5/}.}


\end{thebibliography}







\end{document}